\documentclass[usenatbib]{mn2e}
\usepackage{graphicx}

\newcommand{\text}[1]{\quad\mbox{#1}\quad}

\newcommand{\apj}{ApJ}
\newcommand{\apjs}{ApJS}
\newcommand{\apjl}{ApJ}
\newcommand{\mnras}{MNRAS}

\newcommand{\aap}{A\&A}
\newcommand{\nat}{Nature}
\newcommand{\araa}{ARA\&A}
\newcommand{\apss}{Ap\&SS}
\newcommand{\na}{New Astr.}
\newcommand{\prd}{Phys. Rev. D}
\newcommand{\azh}{Astrono. Rep.}

\title[Model of extended emission of short Gamma-ray Bursts]{Model of extended emission of short Gamma-ray Bursts}

\author[M.V.Barkov \& A.S.~Pozanenko ]{
Maxim V.~Barkov,$^{1,2}$\thanks{
email:~bmv@mpi-hd.mpg.de} and %\footnotemark[1] 
Alexei S.~Pozanenko,$^{2}$\thanks{
email:~apozanen@iki.rssi.ru}\\
$^{1}$Max-Planck-Institut f\"ur Kernphysik, Saupfercheckweg 1, 69117
Heidelberg, Germany\\
$^{2}$Space Research Institute, 84/32 Profsoyuznaya Street, Moscow
117997, Russia}

\begin{document}
\date{Received/Accepted}
\maketitle

%%%%%%%%%%%%%%%%%%%%%%%%%%%%%%%%%%%%%%%%%%%%%%%%%%
\begin{abstract}
%%%%%%%%%%%%%%%%%%%%%%%%%%%%%%%%%%%%%%%%%%%%%%%%%%
{Until now the existence of extended emission is an intriguing
property of short bursts. It is not clear what is the nature of
the extended emission. It might be a rising x-ray afterglow, or it
could be a manifestation of the prolonged activity of a central
engine. We consider short duration gamma-ray bursts,
emphasizing the common properties of  short bursts and short burst
with extended emission. Assuming that the extended emission with
broad dynamic range is a common property of short bursts, we
propose a two jet model which can describe both short main episode
of hard spectra emission, specific for short bursts, and softer
spectra extended emission by different off axis position of
observer. The toy model involves a short duration jet powered by
heating due to $\nu\tilde{\nu}$ annihilation  and long-lived
Blandford-Znajek  jet with significantly narrow opening angle.  
Our proposed model is a plausible mechanism for short duration
burst energization, and can explain both short burst with and
without extended emission within a single class of progenitor.}
\end{abstract}


%%%%%%%%%%%%%%%%%%%%%%%%%%%%%%%%%%%%%%%%%%%%%%%%%%
\section{Introduction}
\label{intro}

It is commonly accepted that Gamma-Ray Bursts (GRB) consist of two
populations - long and short duration bursts. The presence of
separate short Gamma-Ray Burst (SGRB) population  was suggested in
\citep{Mazets1981}, and confirmed by the BATSE experiment
\citep{kouv93}. The nature of SGRB can be merging of compact
companions in close binary systems such as neutron stars (NS) or
NS Black-Hole (BH)   \citep{bnpp84,p86}. Short GRBs have several
distinct phenomenological properties which we briefly discuss.

%\subsection{Short gamma-ray bursts}

SGRBs populate the
short mode of a bi-modal duration distribution. The
duration parameter, $T_{90}$, is defined as continuous interval
comprising the $90\%$ of GRB emission in gamma-ray domain
\citep{koshut96}. It is also accepted that the $T_{90}  < 2 $~s is
a good dividing line for short and long bursts for  mid-energy-range
($30 - 300$ ~keV) experiments e.g. \citep{kouv93}. Later on a
 more complex criterion has been suggested in order to distinguish
phenomenologically short and long GRB
\citep{Donaghy2006astro-ph,Zhang2009ApJ...703.1696Z}.

SGRBs have a harder spectrum than long bursts \citep{kouv93},
and there is no spectral  lag in their light curves in comparison
with long bursts, where light curves of the same GRB in soft channel
lag the light curve in harder channels \citep{Norris1999}.

In a new era of rapidly slewing robotic telescopes, the optical
afterglows of SGRBs has been detected, and in some cases it is
possible to determine a redshift, $z$, of some SGRB sources. It
turn out that most SGRBs have $z < 1$ and equivalent isotropic
energy $E_{iso}$ in the range $10^{48}$ to $10^{51}$~ergs
\citep{Swiftreview}.
% In contrast to long bursts, which occur
% mostly in late-type star-forming galaxies, the host morphology of
%short bursts show no preferable galaxy type
% \citep{BergerHost2009}.
% Finally, there has, today, been no cases of a supernova signature
% being detected, either spectroscopically, or photometrically, in
% the light curve of an afterglow.

Finally,  until now  there has been no cases of a supernova
signature detected, either spectroscopically, or photometrically.

%The analysis of afterglows shows a significantly lower average
%luminosity for short bursts \citep{kann2008SGRBafterglow}.
%On the contrary, it has been shown that the optical (and X-ray) afterglow of short
%and long bursts are very similar \citep{nysewander2009}.

%\subsection{Short gamma-ray bursts with Extended Emission}

\begin{figure}
  \includegraphics[width=1\linewidth]{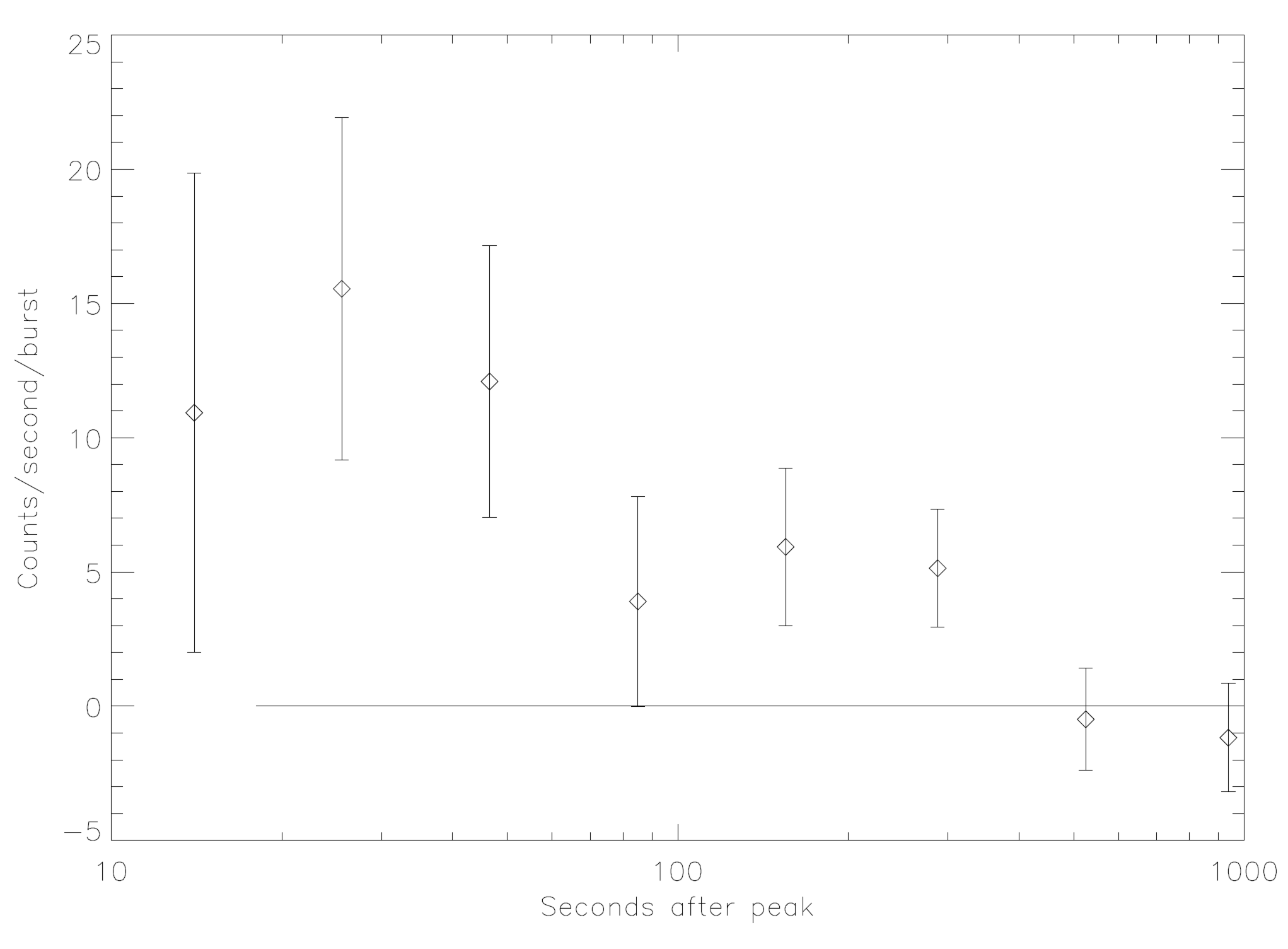}
  \caption{Light curve in 50
- 300~keV energy band for 100 short ($<1$~s), summed, background
subtracted, BATSE bursts after peak alignment. The time interval
corresponding to a peak is not shown, adapted from
\citep{Conn02}.}
  \label{batse_ee}
\end{figure}

One of most intriguing properties of SGRBs is that they exhibit extended emission.
Prompt gamma-ray emission of SGRB consists of a short main episode,
sometimes resolved into substructure of  short pulses, which we
call the Initial Pulse Complex (IPC). The duration, $T_{90}$, of the IPC
 of the short burst is usually less than $\sim 2~s$.
However, in the sum of light curves of many short bursts, aligned
relative to the main peak of the IPC, significant extended
emission (EE) has been observed up to $\sim 100~s$ in different
experiments and energy bands, in particular the BATSE $25-110$~keV
\citep{laz01} and $50-300$~keV \citep{Conn02} (see Figure
\ref{batse_ee}) \citep{Conn02}, Beppo-Sax $40-700$~keV
\citep{Montanari2005}, Konus  $20-750$~keV \citep{FrederiksEE},
and SPI-ACS $>80$~keV \citep{Minaev09, Minaev2010a}. Indeed, the
$T_{90}$ parameter of each short burst to be summed is less than
$\sim 2~s$. The intensity of the EE seems to depend on the energy
band being smaller with increasing the energy band, and fluence
ratio of the EE episode and the IPC may be less than 1/100.

\begin{figure}
  \includegraphics[width=1\linewidth]{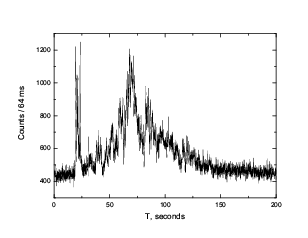}
  \caption{Light curve of GRB060614 in 15 - 150~keV energy band as recorded with BAT/Swift, time bin is 64 ms.}
  \label{GRB060614_BAT_lc}
\end{figure}

The EE is softer than the  IPC, however it is not clear yet if this EE is a rising X-ray afterglow
or manifestation of prolonged activity of the central engine of the
burst source. Also, based only on the statistical investigation of
a large amount of short GRBs, one cannot say for sure that each
particular burst has extended emission.
%However, since there is
%no additional population of short bursts found to date, one can
%suppose that extended emission is a  common feature of SGRB, and
%the extended emission has a different intensity in comparison with the IPC.

Extended emission has actually been observed in individual light
curves of some SGRB, confirmed with KONUS \citep{FrederiksEE},
HETE-2  \citep{Villasenor2005Nature}, Swift
\citep{Barthelmy2005Nature}, BATSE \citep{Norris06},
SPI-ACS/INTEGRAL \citep{Minaev2010a}. Despite the $T_{90}$  of
some bursts being less than 2 seconds, those the EE are
significantly detected after re-binning original light curve onto
larger time scales.  In all cases above, the EE has a peak flux
and fluence much smaller, than analogues parameters of the IPC.

Among the long duration BATSE bursts, several GRBs with
phenomenological features resembling short bursts were found,
where IPC does not possess  any spectral lag and there is a tail
of softer emission with durations up to $100~s$ \citep{Norris06}.
The visually estimated amount of this burst type is about $ 2\%$
of the total amount of BATSE bursts. While the peak flux of the EE
is still 10-30 times less than the peak flux  of the IPC, the
fluence of the EE is compatible with the fluence of the IPC.

Finally, the ultimate example of an apparently long burst ($T_{90}
= 102~s$, see Figure \ref{GRB060614_BAT_lc}) with all signatures
of short  bursts is Swift GRB~060614 \citep{GehrelsNature2006}. In
addition to the absence of spectral lag in the whole light curve
(both in the IPC and the EE), the host galaxy was identified at
z=0.125 and deep and temporally dense follow-up optical
observation did not reveal signature of a supernova, placing the
brightness of any possible supernova associated with GRB 060614
unreasonably faint $M_{R}> -13.6$ \citep{Fynbo2006Natur,
Gal-Yam2006Natur, DellaValle2006Natur}.

It is evident that the intense EE of SGRBs was observed in a few
cases, whilst the less intense EE can be observed  more frequently.
Using BAT/Swift, which is more sensitive to the EE of short bursts, it
was shown that among Swift SGRBs that $\sim 25\%$ of short
bursts have EE  \citep{Norris1101}. It is also evident that
weak EE are observed in the \textit{majority} of short bursts because sum
of the light curves possess EE, whilst in individual light curves,
the EE is below the detection level.

\begin{figure}
  \includegraphics[width=0.8\linewidth]{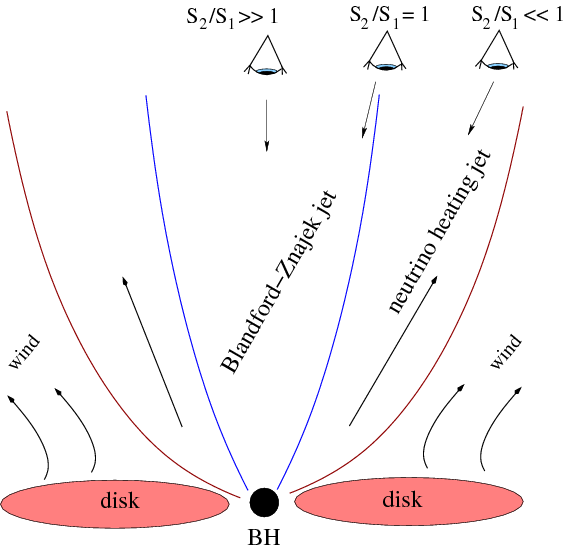}
  \caption{A scheme of a two component jet model.
  Observer will register different fluence of short episode of initial
  emission and extended emission depending on the angle of view against axis of the jets.
%  The frequency distribution of flunce ratio will be proportional
%  to the ratio of solid angles of the Blandford-Znajek and neutrino heating jets.
}
  \label{sketch}
\end{figure}

Thus one could suggest that for short duration bursts (or the
phenomenon responsible for what we mean as short duration bursts)
have a distinctive feature, such as extended emission with very broad
dynamic range of  flux and fluence of  the EE.
We suggest that
EE is an inherent property of short bursts, and an observable
property of the EE can be explained by the different angular position
of the observer with respect to the axis of the coaxial jets (see Figure
\ref{sketch}).

%%%%%%%%%%%%%%%%%%%%%%%%%%%%%%%%%%%%%%%%%%%%%%%%%%%%

%%%%%%%%%%%%%%%%%%%%%%%%%%%%%%%%%%%%%%%%%%%%%%%%%%%%

%%%%%%%%%%%%%%%%%%%%%%%%%%%%%%%%%%%%%%%%%%%%%%%%%%
\section{The model}
\label{model}

In our model, we assume the merging of two NSs or a NS+BH system which can
lead to formation of a fast-rotating black-hole ($a=J/M_{BH}^2$,
where $a$ is dimensionless BH spin parameter). Simulations of
the merging of the BH+BH or BH+NS give values of final $a$ as
$0.3\le a\le0.65$ \citep{bbc08,rj10}, and in our
calculations, we will assume $a=0.5$, the formation of relatively
massive $M_d\sim 0.001-0.2 M_{\odot}$, and compact $R_d\sim
10^7~cm $ accretion disk \citep{lr07}. Such a disk is optically
thick and photon cooling is not efficient and in the case of
ineffective neutrino cooling, such a disk also becomes geometrically thick.

To explain the extended emission of SGRBs, we suggest a
two-component model with neutrino heating \citep{w93} and an
electromagnetic Blandford-Znajek (BZ) mechanism
\citep{BZ77,lbw00,myks04,BK08b,rgb11}.  A short main episode (IPC)
of SGRB is the result of a fast accretion period \citep{pwf99},
which launches  a neutrino powered jet. After a few seconds,
neutrino heating becomes ineffective, however, the lower accretion
rate can keep the central engine active for a longer time due to
the BZ mechanism \citep{BK10}. {The duration of the BZ powered jet
will depend on the mass in accretion disk.}

The accretion time of main mass of the compact disk with radius $R_d\sim 10^7 $
cm is short $0.1-1$ s \citep{vpo01}. Following  \cite{mpq08}, the
accretion rate   can be estimated as:
\begin{equation}
\dot{M}_d\approx fM_d/t_{visc},
\label{dmdt}
\end{equation}
where $t_{visc}=R_d^2/\nu$ is viscosity time scale and $\nu$ is the viscosity,
and the factor $f\approx 1.6$. For the viscosity we have used an $\alpha$-prescription
\citep{s72,ss73}, $ \nu= \alpha c_s H,$
where $c_s=(P/\rho)^{1/2}$ is the isothermal sound speed and $H$ is
half-thickness of the disc. The initial viscosity time scale is:
\begin{equation}
t_{visc,0}\approx 0.02
\alpha^{-1}_{-1}M_{0.5}^{-1/2}R_{d,7}^{3/2}\left(\frac{H}{R_d}\right)^{-2} \mbox{
s,}
\label{tvisc}
\end{equation}
where $M_{0.5}=M/10^{0.5} M_{\odot}$ is mass of the BH, $\alpha_{-1}=\alpha/0.1$ is the
standard dimensionless viscosity,
$R_{d,7}=R_d/10^7$ cm is the radius of the accretion disk.

The accretion disk will be thick and radiatively inefficient for
neutrino cooling if it is larger $R_d>300 R_g$ or $t>t_{thick}\sim
2 \alpha_{-1} $ s \citep{CB07}, where $R_g=GM_{BH}/c^2\approx 5\times10^5
M_{0.5}~\mbox{cm}$. For the calculation of the extended emission phase, we can use
such an approximation, whilst $t>t_{thick}$, the radioactively inefficient advective
flux approximation is applicable. We follow the work \cite{bb99}, and assume that
only  a fraction $\sim (R/R_d)^p$ of available material is
accreted onto the  BH. The rest of the mass will be lost to an
outflow, and the parameter, 'p', can vary from 0 (no wind) up to 1
(powerful wind). The self-similar solution for accretion rate was
obtained by \cite{mpq08}. This solution gives us an accretion rate:
$$
\dot{M}_{in}\approx
1.6\frac{M_{d,0}}{t_{visc,0}}\left(\frac{R_{ms}}{R_{d,0}}\right)^{p}\times
$$
\begin{equation}
\left[1+4.8(1-C)\left(\frac{t}{t_{visc,0}}\right)\right]^{-[1+3(1+2p/3)(1-C)]/[
3(1-C)]},
\label{dmdtw0}
\end{equation}
and disk radius
\begin{equation}
r_d\approx
R_{d,0}\left[1+4.8(1-C)\left(\frac{t}{t_{visc,0}}\right)\right]^{2/3},
\label{rw0}
\end{equation}
where $M_{d,0}$, $R_{d,0}$ and $t_{visc,0}$ are initial mass,
radius and viscous time of the disk, and $C=2p/2p+1$. The $R_{ms}$ is  radius of
the marginally stable orbit (MSO)  \citep{bpt72}.

For the accretion of a thick disk, the maximal luminosity due to
BZ mechanism can be estimated following the papers
\citep{KB10,B10}\textbf{ where used effect of magnetic field grows
and a strong large scale magnetic field formation in the vicinity
of the black hole horizon \citep{BKR74,BKR76}. }

{\bf The pressure of the magnetic field can be a
fraction, $1/\beta$ (where $\beta \equiv 8\pi P_g/B^2 $), of the
gas pressure in the disk at MSO, from MSO magnetic field accretes
to the BH horizon.} In such a way, the luminosity of the BZ
mechanism becomes a weak function of BH spin parameter, $a$, (if
$0.5\leq a \leq 1$) and can be estimated as:
\begin{equation}
L_{BZ}\approx \frac{0.05}{\alpha_{-1}\beta_1} \dot{M}_{in}c^2\approx
10^{48} \alpha_{-1}^{-1}\beta_1^{-1} \dot{M}_{in,-5} \mbox{ erg s}^{-1},
\label{lbz}
\end{equation}
where $\beta_1=\beta/10$.

The main source of neutrino heating is the neutrino annihilation
reaction $\nu\tilde{\nu} \rightarrow e^+e^-$. The heating rate can
be described by  \citep{zb10}:
\begin{equation}
L_{\nu\tilde{\nu}}\approx 3\times 10^{50}
X^{-4.7} M_{BH,0.5}^{-3/2} \dot{M}_{in,0}^{9/4} \mbox{ ergs s}^{-1},
\label{enunu}
\end{equation}
where $X\equiv R_{ms}/{4R_g} $.
This formula is valid when the accretion rate is higher than $\sim
0.05 \alpha_{-1}^{5/3}  M_{\odot} $ s$^{-1}$ \citep{CB07}, this
critical value of the accretion rate is a function of $a$. As the
accretion rate becomes lower, the efficiency  of neutrino heating
drops dramatically and becomes negligible.

In our model, the distribution of SGRBs in the intensity of the
extended emission is a selection effect  of different angular
position of the observer (see Figure \ref{sketch}) with respect to the axis of
coaxial jets and a dispersion of opening angles of the neutrino
and BZ jets.
 The neutrino powered jet has the opening angle
$\theta_{\nu\tilde{\nu}}\sim 0.1$
 \citep{ajm05,hkts10} and can  be significantly wider than
opening angle of the jet powered by the BZ mechanism $\theta_{BZ} \sim
1/\Gamma$ \citep{KVKB09}.

One can estimate the ratio of jets opening angles using GRB
060614. As discussed above, this unique burst has brightest EE,
i.e. peak fluxes nearly the same in the IPC and the EE (Figure
\ref{GRB060614_BAT_lc}). One can suggest that GRB 060614 was
observed close to the axis of the coaxial jets. Then we can write
following relation:
 \begin{equation}
\frac{ L_{BZ} }{  L_{\nu\tilde{\nu}}
}\left(\frac{\theta_{\nu\tilde{\nu}}}{\theta_{BZ} }\right)^2
=\frac{F_{2,max}}{F_{1,max}} \sim 1
 \label{peak_flux_ratio}
\end{equation}
%$
%\frac{ L_{BZ} }{  L_{\nu\tilde{\nu}}
%}\left(\frac{\theta_{\nu\tilde{\nu}}}{\theta_{BZ} }\right)^2
%=\frac{F_{2,max}}{F_{1,max}} \sim 1
%$
where $F_{1,max}$ and $F_{2,max}$ are peak fluxes of the IPC and EE
episodes. Using equations (\ref{lbz},\ref{enunu}), we obtain $\theta_{BZ}/
\theta_{\nu\tilde{\nu}} \sim 0.1$. Otherwise, one can estimate the
same parameter of $\theta_{BZ}/ \theta_{\nu\tilde{\nu}}$ using
fluence ratio of the IPC with duration, $t_{\nu\tilde{\nu}}$, of
about 5 s and fluence $S_{1}= (16.9 \pm 0.2)\times
10^{-6}$~erg~cm$^{-2}$, and the EE episode with duration $t_{BZ}$
of about 100 s and fluence $S_{2} = (3.3 \pm 0.1)\times
10^{-6}$~erg~cm$^{-2} $ (see \cite{GehrelsNature2006}):
 \begin{equation}
\frac{L_{BZ}}{L_{\nu\tilde{\nu}}}
\left(\frac{\theta_{\nu\tilde{\nu}}}{
 \theta_{BZ}}\right)^2  \frac{t_{BZ}
}{ t_{\nu\tilde{\nu}}} = \frac{S_{2}}{S_{1}} \sim 5,
 \label{flunce_ratio_5}
\end{equation}
which leads to the comparable value of $\theta_{BZ}/
\theta_{\nu\tilde{\nu}} \sim 0.1$.

Frequency detection of the EE in the individual light curves of
short GRB  is roughly proportional to the ratio of solid angles of
the two jets. The frequency ratio can be estimated from 2\% of all
burst of BATSE \citep{Norris06}, and up to 25\% of short bursts
registered by Swift \citep{Norris1101}. Hence we have $
\left(\frac{\theta_{BZ}}{ \theta_{\nu\tilde{\nu}}}\right)^2  =
\frac{1}{50} \div \frac{1}{4} $ and $\theta_{BZ} /
\theta_{\nu\tilde{\nu}} = 0.15  \div 0.5$.  One can assume that
the opening angle ratio of Blandford-Znajek and neutrino powered
jets is between 0.5 and 0.1. {(In above estimations, we ignore any
possible angular distribution of energy releases and
$\Gamma$-factors of emitting particles in the jets. We also
implicitly assume in
eqs.(\ref{peak_flux_ratio},\ref{flunce_ratio_5}) the conversion
factors to the gamma-ray emission are equal in both jets.)}

%%%%%%%%%%%%%%%%%%%%%%%%%%%%%%%%%%%%%%%%%%%%%%%%%%%%
\begin{figure}
  \includegraphics[width=0.8\linewidth]{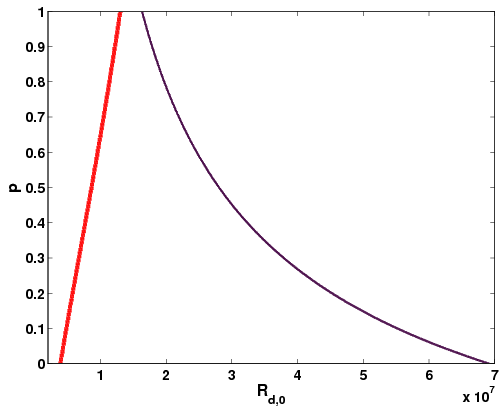}
  \caption{First episode of short emission and the extended emission is observable if initial parameters 'p' and $R_{d,0}$ is in between the lines.
  We take the initial mass of the accretion disk is equal to $M_{d,0}=0.1 M_{\odot}$,
  $a=0.5$ and duration of the IPC and the extended emission equals $t_{\nu\tilde{\nu}}=1~s$, and $t_{BZ}=100~s$.
  The area to the right of thick line corresponds to  $L_{\rm BZ}\geq 3\times 10^{-4}L_{\nu\tilde{\nu}}$.
  The area to  the left of thin line represents the initial accretion rate $\dot{M}_{in}>0.05 M_{\odot}$ s$^{-1}$  }
  \label{pr_plot}
\end{figure}

%%%%%%%%%%%%%%%%%%%%%%%%%%%%%%%%%%%%%%%%%%%%%%%%%%%%

%ssssssssssssssssssss
\section{Discussion}
%ssssssssssssssssssss

The gamma-factors of the two jets can be also different, with
$\Gamma_{\nu\tilde{\nu}} \sim  500$  predicted from modelling with
relativistic hydrodynamic codes for outflows \citep{ajm05}. {
The interpretation of observations with a simplified one-zone model
suggests extremely high Lorentz factor of $\Gamma > 1200$ for the IPC
of the short GRB 090510 \citep{Ackermann_GRB090510_2010ApJ}; see also
\cite{Corsi_GRB090510_2010ApJ}. Nevertheless, two-zone models of
GeV photon production gives a significantly lower limitation on
the Lorentz factor of the jets of $\Gamma > 200-400$ \citep{zfp11}.}

\textbf{ $\Gamma_{BZ}$ can be estimated  from the value of opening
angle $\theta_{BZ} \sim 1/\Gamma_{BZ}$ which is specific for
BZ-jet \citep{KVKB09}, and ratio
$\theta_{BZ}/\theta_{\nu\tilde{\nu}}<0.5$. If
$\theta_{\nu\tilde{\nu}} \sim 0.1$ we can obtain the lower limit
of $\Gamma_{BZ} > 20$, which is compatible with numerical
calculations of BZ-jet parameters at the time the jet is started.
In the course of BZ-jet development, the $\Gamma_{BZ}$  increases
with time \citep{KVKB09}. Then the BZ jet leaves the zone of
collimating wind, and the outflow can get additional acceleration
due to the propagation of a rarefaction wave across the entire jet
\citep{tnm10,kvk10}.  This additional boosting will not change
significantly the BZ-jet opening angle while the $\Gamma_{BZ}$
could reach  $100-300$.}

The spectra of the IPC of SGRBs are always harder than the spectra
of the extended emission and it could be explained by
$\Gamma_{\nu\tilde{\nu}}> \Gamma_{BZ}$ {\bf (see however
\cite{bdm05})}. Another issue is an absence of the spectral lag.
The spectral lag is unavoidable if the emitting  particles are
moving radially
\citep{Fenimore1996ApJ...473..998F,Sari1997ApJ...485..270S,Norris2002ApJ}.
The absence of  the spectral  lag in the IPC may be explained  by
large value of $\Gamma$ \citep{Norris06}, although, it is not
clear whether the lag is present in  the EE. In a few cases, where
the lag could be accurately measured, it is negligible within
statistical errors \citep{Norris06,GehrelsNature2006}. {In
general, more investigations of phenomenology and nature of the
spectral lag are necessary. }

The IPC of GRB 060614 falls far off \citep{GehrelsNature2006} the
 Amati relationship for long GRBs \citep{Amati_correlation2002A&A,Amati2010Fermi}.
However, the $E_{p}$, $E_{iso}$ parameters of the whole burst are
consistent, within 2 sigma level, with the Amati relationship for long
GRBs \citep{Mangano2007GRB060614}. One would observe the GRB
060614 off-axis of the BZ-jet, such could not detect the EE, but only IPC.
In this case, eventually GRB 060614 would be an outlier of the Amati
relationship as we see it for other short duration bursts
\citep{Amati2010Fermi}.

Finally let us estimate physical parameters when the EE is
observable. Based on the equation (\ref{flunce_ratio_5}) and typical values
obtained for extended emission registered for sum of the short
duration bursts, i.e. fluence ratio $S_{2}/S_{1} \sim 1 $,
$t_{\nu\tilde{\nu}} \sim 1~s$, $t_{BZ} \sim 100~s$ and mean value
of $\theta_{BZ} / \theta_{\nu\tilde{\nu}} \sim 0.2$, we can infer
that the extended emission can be detected if
$L_{BZ}/L_{\nu\tilde{\nu}} > 3\times 10^{-4}$.

Using eq.(\ref{dmdtw0}) at time  $t=0$, we can draw  in the
Figure \ref{pr_plot} a (thin) line which shows the limit
$max(\dot{M}_{in})=0.05 M_{\odot}$ s$^{-1}$.  Thick  line
represents the limit $L_{BZ}/L_{\nu\tilde{\nu}}=3\times10^{-4}$,
and the area to the right of this line implies that there is enough energy in
the BZ jet to be detected. The region between the lines provides a
region in which both IPC and extended emission could be detected.
Indeed in the region on the left of the thick line
($L_{BZ}/L_{\nu\tilde{\nu}}<3\times10^{-4}$), the accretion rate is ether
too slow to start BZ jet or to provide enough energy in the
EE.

On the other hand, in the region to the right of thin line, the
initial short period of a fast accretion rate providing a neutrino
powered jet is absent. However, the accretion rate can be
sufficient to start the BZ jet. Therefore one can expect the
existence of rare type of long duration bursts which look like
only the extended emission of such bursts as GRB 060614. {The
observable features of the burst are the absence of a supernova,
negligible  spectral lag, and a possible precursor corresponding
to a failed neutrino jet.} It is difficult to estimate a frequency
of the occurrence of such bursts as it will depend on the initial
accretion rate distribution. However, one can remember that $\sim
10\%$  of long BATSE bursts \citep{Hakkila_lag_2007ApJS} actually
possess zero spectral lag (see also \citep{Norris2002ApJ}).
Indeed, some of them are short bursts with observable extended
emission, as found by \citep{Norris06}, whilst some of them could
be  only BZ powered bursts. Recent spectral lag investigation of
naturally long burst of the BAT/Swift experiment gives the same
$\sim$ 10\% of burst which have negligible lag within statistical
error \citep{Ukwatta2010ApJ...711.1073U}.

The main source of the magnetic field which is necessary to start
the BZ jet can be a seed magnetic field of the merging
companions. The variability of magnetic flux can be a manifestation
of a magnetic dynamo in the accretion disk and the accretion of
magnetic flux onto the BH \citep{BB11,rgb11}. The accretion of magnetic
flux with different polarity leads to the variability  of magnetic
flux on the BH horizon and thus a variability of BZ luminosity as well.

The  magnetic field generated due to a dynamo can have a pressure,
{ which can be} comparable with the pressure of matter
\citep{BKL07,RL08,I08,Lovelace_2009ApJ...701..885L}.  The pressure
of the gas in the disk is a function of its thickness, $P\propto
(H/R)^2$, and hence $B\propto (H/R)$, whilst the energy release
will be $L_{BZ} \propto B^2 \propto (H/R)^2$. On the time scale of
$t<t_{thick}$, the magnetic dynamo can generate significantly
weaker fields than in the thick disk and luminosity of BZ is
suppressed. In the time scale of $t>t_{thick}$ which may be of few
seconds the disk becomes thicker, the magnetic dynamo becomes more
efficient and provides an increased BZ-jet luminosity. The
increase of the luminosity may explain the hiatus between the IPC
and the extended emission observed in some short burst with EE
\citep{Norris06,GehrelsNature2006}.

{ Finally, the same mechanism involving two jets can be
responsible not only for GRB based on NS-NS or NS-BH merging.
Indeed the merging of BH or NS and white dwarf
\citep{Fryer_BH_WD_1999ApJ} or BH and a core of Wolf-Rayet star
\citep{FW98} can rise a long duration burst and  exhibit a long
X-ray afterglow \citep{BK10} with appropriate time scaling of the
main episode powered by a neutrino jet and extended emission,
powered by BZ jet. The time scaling parameter will be dependent on
the initial linear size (specific angular momentum) of the
accretion disk, and in the case of a BH and white dwarf merging,
may be $\sim 10$.}

%%%%%%%%%%%%%%%%%%%%%%%%%%%%%%%%%%%%%%%%%%%%%%%%%%%%%%%%%%%%%%%%%
\section*{Acknowledgments}
We are grateful to G.~S. Bisnovatyi-Kogan for useful discussions.
The work was supported by the Origin and Evolution of Stars and
Galaxies Program of Russian Academy of Sciences.
%%%%%%%%%%%%%%%%%%%%%%%%%%%%%%%%%%%%%%%%%%%%%%%%%%%%%%%%%%%%%%%%%

%\bibliographystyle{mn2e} % style mn2e.bst
%\bibliography{barkov_base_1m}

\end{document}